# Booster Synchrotron Frequency Below Transition

Xi Yang, James MacLachlan, Rene Padilla, and C. Ankenbrandt
*Fermi National Accelerator Laboratory*
Box 500, Batavia IL 60510

**Abstract**

The dipole mode synchrotron frequency is a basic beam parameter; it and a few similarly basic quantities measured at small time intervals serve to characterize the longitudinal beam dynamics throughout the acceleration cycle. The effective accelerating voltage, in conjunction with the amount of rf voltage required for the acceleration, is important for the estimate of the beam energy loss per turn. The dipole mode frequency can be used to obtain the effective accelerating rf voltage, providing that it can be measured precisely. The synchrotron frequency measured from the synchrotron phase detector signal (SPD) generally agrees well with calculation, and it can be applied for such purposes as inferring the effective rf voltage.

## Introduction

It is important to have more than one method for estimating the effective rf voltage in the Booster. First, the effective accelerating rf voltage ($V\times sin(\phi_s)$) can be obtained from the direct measurements of the accelerating rf voltage ($V$) and the synchronous phase ($\phi_s$). Secondly, it can be calculated using eq. (1) from the synchrotron frequency ($f_s$) measured by the SPD[1]:

$$f_s = (f_0)\times\sqrt{heV|\eta_0 \cos(\phi_s)| \Big/ (2\pi\beta^2 E)} \tag{1}$$

Here $h$ is the harmonic number of the rf frequency, 84 in the Booster. $f_0$ is the revolution frequency, $\eta_0$ is the phase slip factor, $\beta$ is the Lorentz's relativistic factor, and $E$ is the total energy.[2]

The agreement between the calculated and the measured synchrotron frequency becomes important for deciding whether or not the measured synchrotron frequency from the SPD signal can be trusted in estimating the effective accelerating rf voltage.

## Experimental Result

The signal to noise ratio from the SPD is improved by connecting its output through a low-pass filter with a cutoff frequency of 40 KHz because the bandwidth of the SPD passes much other information besides the phase of the rf fundamental component of the beam current. The synchrotron frequency was measured at 1 ms intervals through the period of injection (0 ms) to transition (17 ms) for the extracted beam intensity of $4.3\times10^{12}$ protons. The synchrotron frequency was obtained from the modulation on the SPD signal, as shown in Figs. 1(a) – 1(d) for the examples of 0 ms, 1 ms, 10 ms and 15 ms. The synchrotron frequency measured from the SPD signal is shown as the black curve in Fig. 2.

## Calculation of Synchrotron Frequency

The effective rf voltage (RFSUM) and the synchronous phase were measured on separate cycles with similar extracted beam intensity of $4.3\times10^{12}$ protons. RFSUM is shown as the blue curve and the synchronous phase is shown as the black curve in Fig. 3(a). The Lorentz relativistic factor ($\beta$), the phase slip factor ($\eta_0$), the revolution frequency ($f_0$), and the total energy ($E$) of a proton change during the acceleration in a Booster cycle. $\beta$ is shown as the black curve and $\eta_0$ is shown as the blue curve in Fig. 3(b) for a Booster cycle, and $f_0$ is shown as the black curve and $E$ is shown as the blue curve in Fig. 3(c). Thus, all the parameters required by eq. (1) for the synchrotron frequency calculation are known; the results are shown as the red curve in Fig. 2(a).

## Conclusion

The synchrotron frequency was obtained in the period of injection to transition both by direct measurement from the SPD signal and by calculation using eq. (1). The measurement agrees with the calculation well except during the first two ms after injection. Thus the synchrotron frequency can be applied for the estimate of the effective accelerating rf voltage in conjunction with the direct RFSUM and synchronous phase measurements. The discrepancies between the measurement and the calculation at the first two ms after injection need further investigation, but macroparticle modeling agrees more closely with the measurements, and is shown as the green curve in Fig. 2. Also, the synchrotron frequency measurement after transition needs more effort because the dipole mode signal is swamped by the excitation of bunch length oscillation at transition.

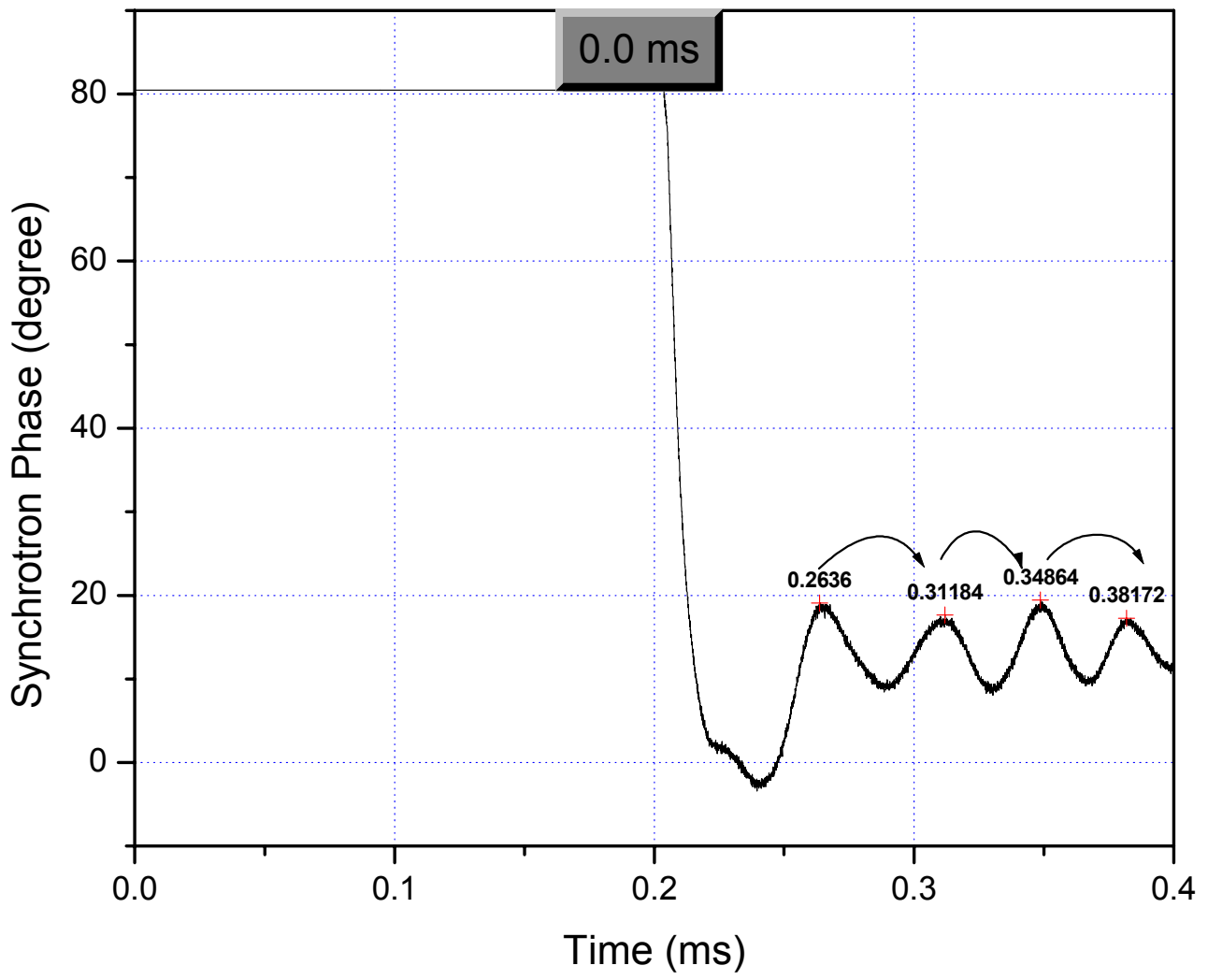
Fig. 1(a)

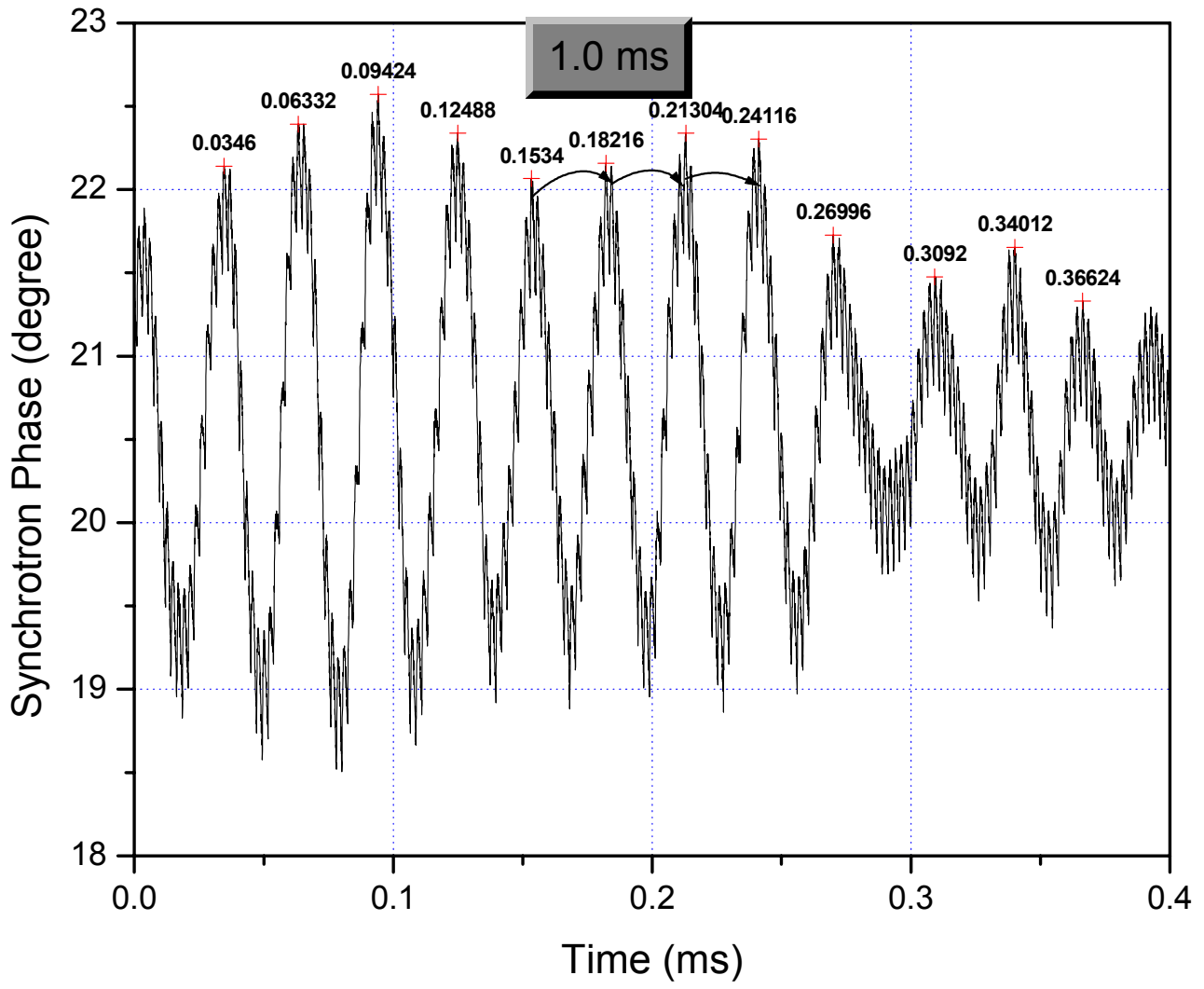
Fig. 1(b)

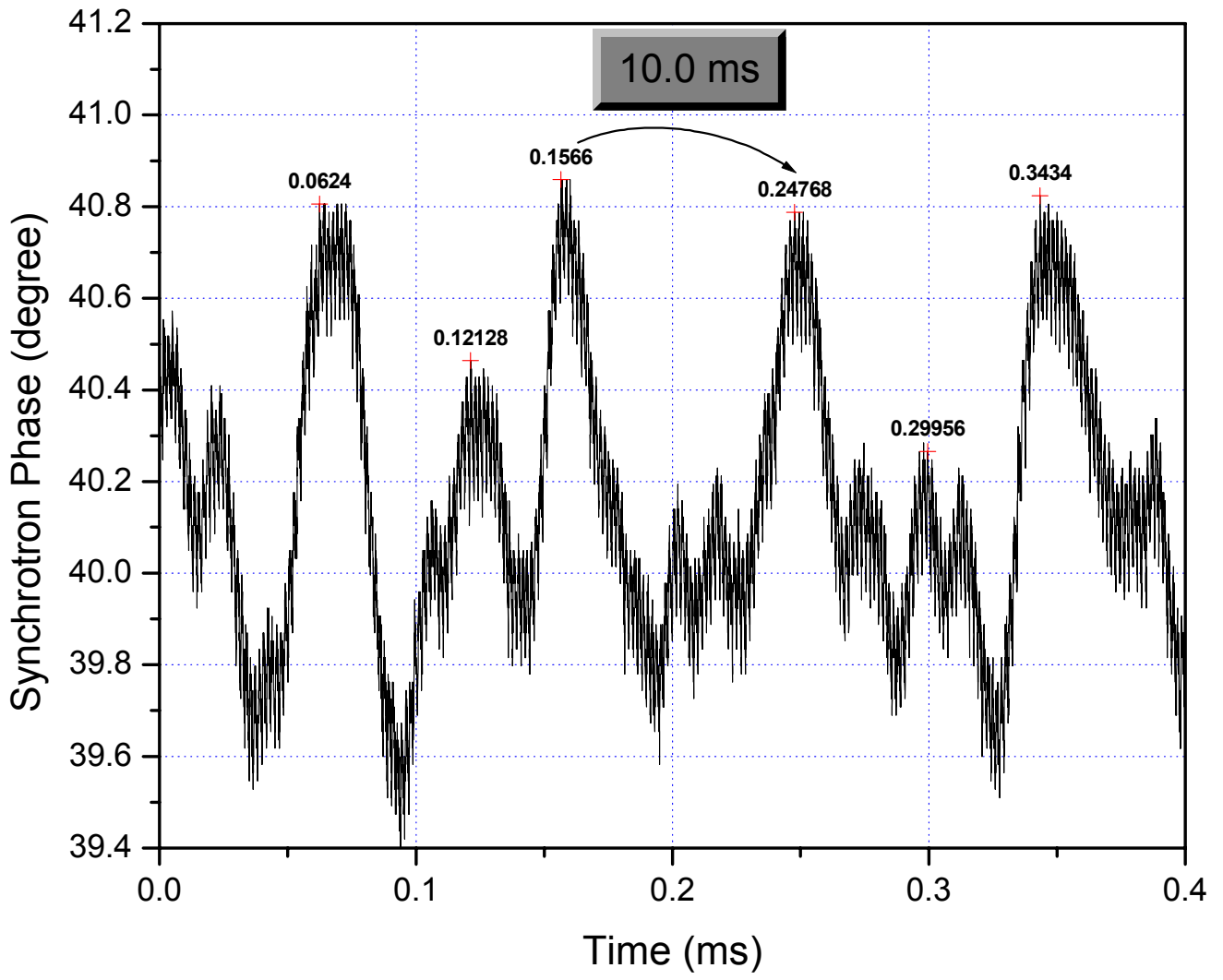

Fig. 1(c)

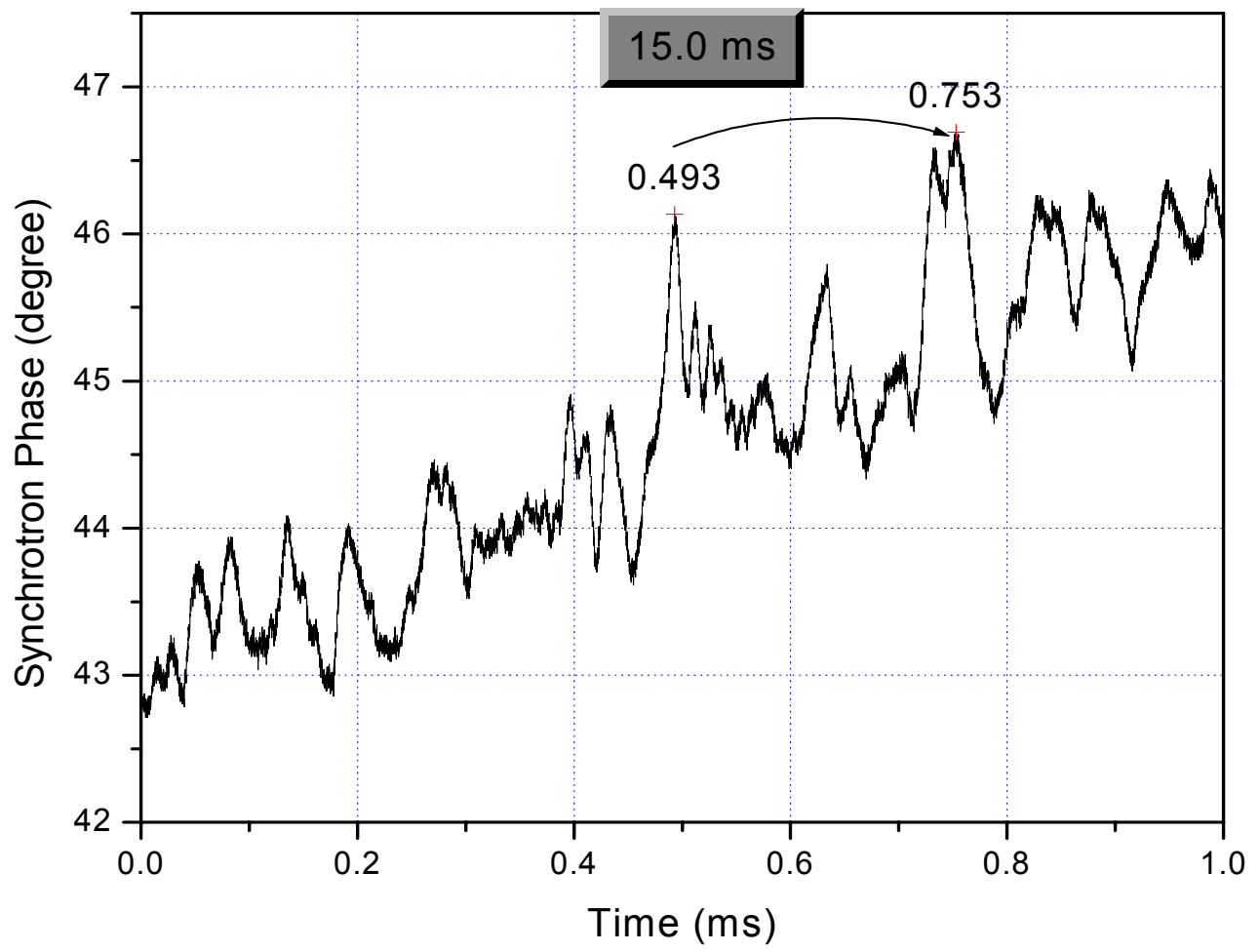

Fig. 1(d)

Fig. 1 The SPD signal obtained at the extracted beam intensity of $4.3\times10^{12}$ protons. The modulation on the SPD signal was used for the measurement of the synchrotron frequency. The trigger time for the measurement was

Fig. 1(a) 0.0 ms at injection.

Fig. 1(b) 1.0 ms after injection.

Fig. 1(c) 10.0 ms after injection.

Fig. 1(d) 15.0 ms after injection.

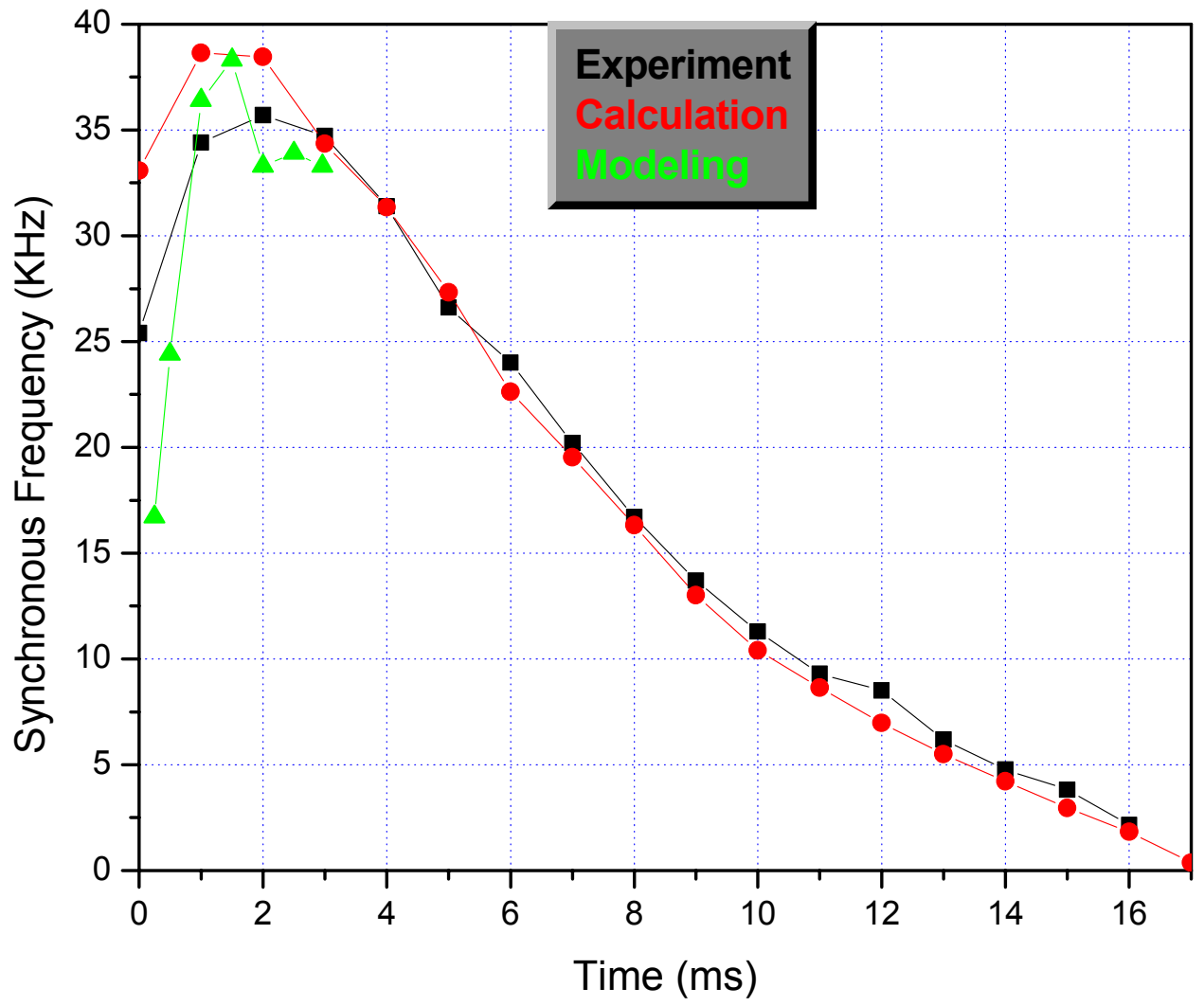

Fig. 2

Fig. 2 the measured synchrotron frequencies from the SPD signal are shown as the black curve, the calculated synchrotron frequencies are shown as the red curve, and the synchrotron frequencies at the first 3 ms after injection from macroparticle modeling are shown as the green curve.

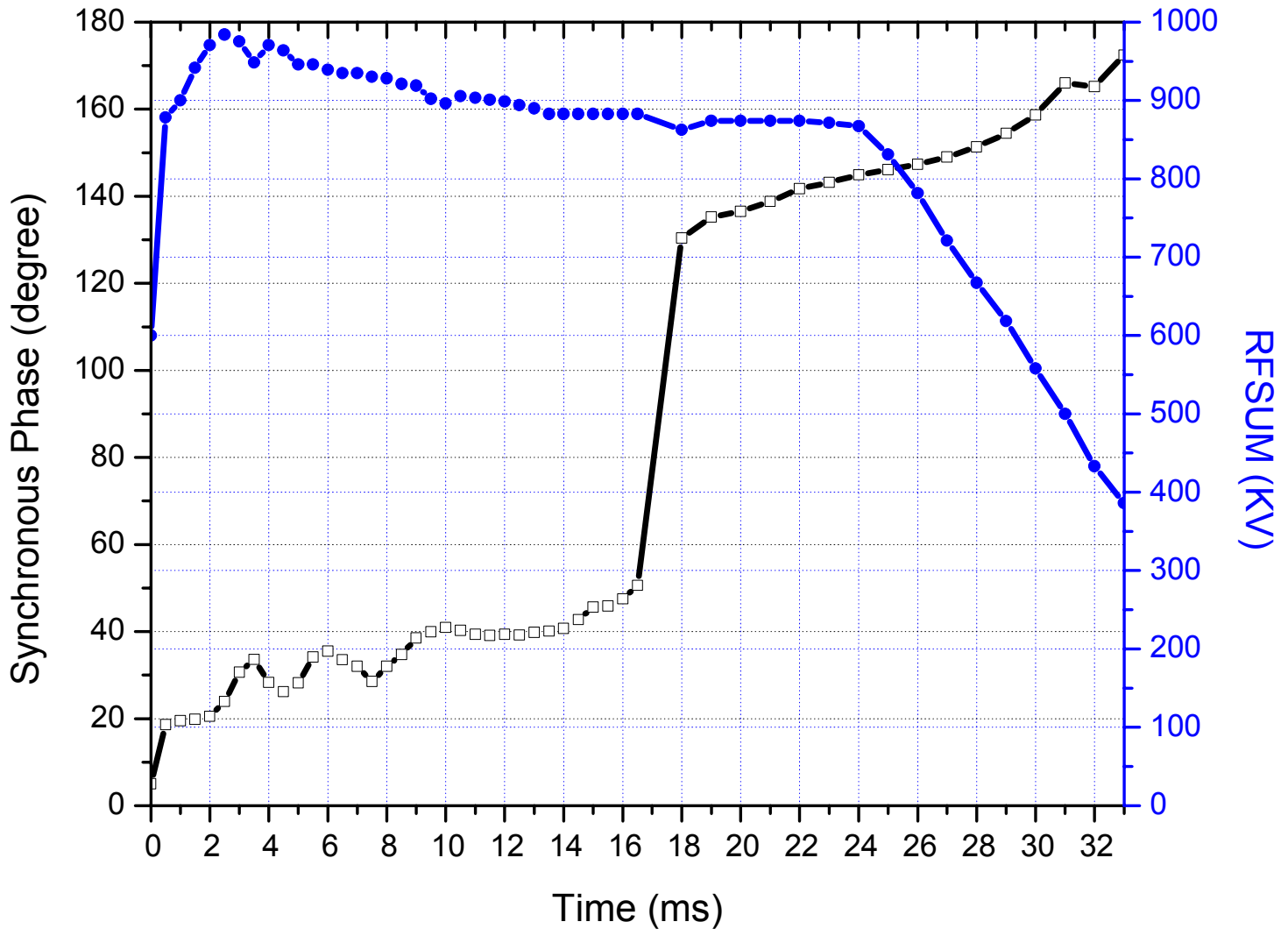

Fig 3(a)

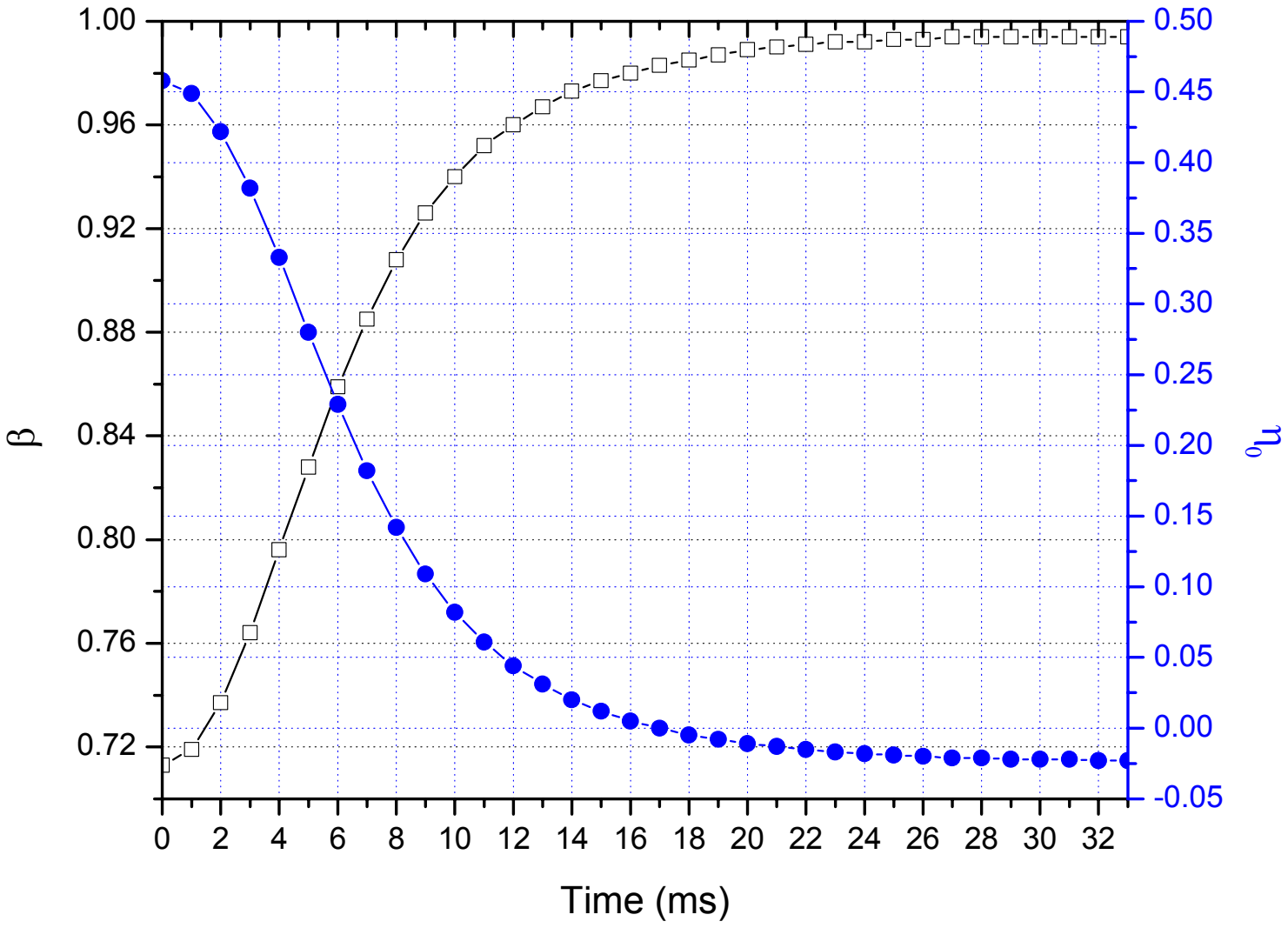

Fig. 3(b)

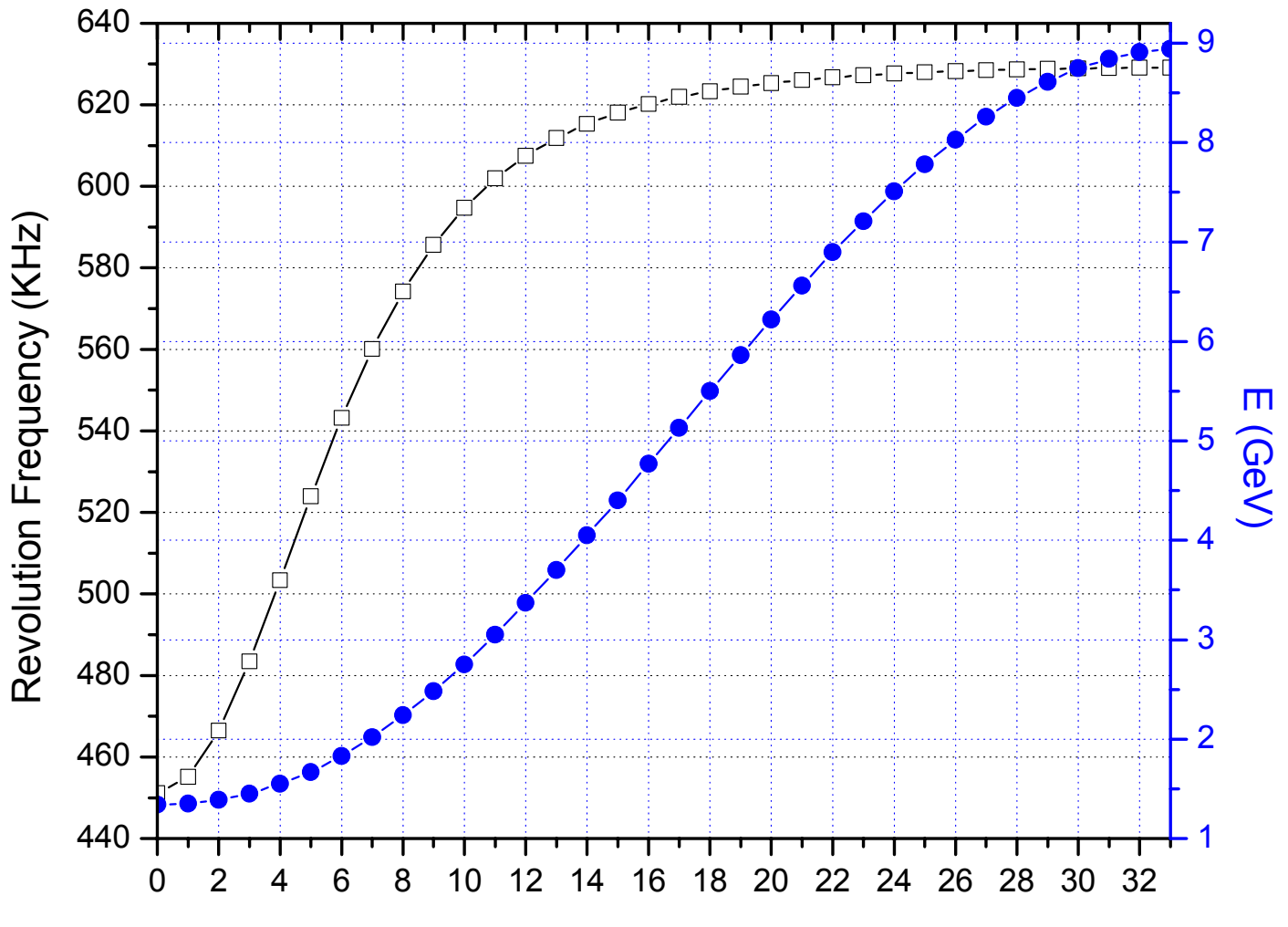

Fig. 3(c)

Fig. 3(a) the synchronous phase and RFSUM measured at the extracted beam intensity of $4.3\times10^{12}$ protons.

Fig. 3(b) the calculated Lorentz's relativistic factor ($\beta$) for a Booster cycle is shown as the black curve, the calculated phase slip factor ($\eta_0$) is shown as the blue curve.

Fig. 3(c) the revolution frequency ($f_0$) for a Booster cycle is shown as the black curve, and the total energy ($E$) of a proton is shown as the blue curve.